\documentclass[12pt]{article}

\input epsf.tex
\usepackage{amsmath,bbm}
\usepackage{amssymb}
\usepackage{graphicx}

\begin{document}
\font\vmath=msbm10 at 12pt

\newcommand{\vmi}[1]{\mathbbm{#1}}
\def\sp{\phantom{a}}

\begin{titlepage}

\font\csc=cmcsc10 scaled\magstep1
{\baselineskip=14pt
 \rightline{
 \vbox{\hbox{RIKEN-TH-56}}}}
{\baselineskip=14pt
\rightline{
 \vbox{\hbox{hep-th/0511031}
    }}}

{\bf
       \vspace{2cm}
       \begin{center}
         {Quantization of fields based on Generalized Uncertainty Principle}     
       \end{center}
}

\vspace{1cm}

\begin{center}
{Toshihiro Matsuo\footnote{\tt tmatsuo@riken.jp} and
SHIBUSA Yuuichirou \footnote{\tt shibusa@riken.jp}}

\vspace{0.5cm}
{\it
Theoretical Physics Laboratory,\\
The Institute of Physical and Chemical Research (RIKEN),\\
Wako, Saitama, 351-0198, Japan
}

\end{center}
\vspace{2cm}
\begin{abstract}
We construct a quantum theory of free scalar field in $1+1$ dimensions based on the deformed Heisenberg algebra $[\hat{x},\hat{p}]=i \hbar(1+\beta \hat{p}^2)$ where $\beta$ is a deformation parameter.
Both canonical and path integral formalism are employed.
A higher dimensional extension is easily performed in the path integral formalism.
\end{abstract}

\end{titlepage}

\baselineskip=16pt
\setcounter{footnote}{0}

\section{Introduction}

Physics in extremely high energy regions is particularly of interest to particle physics. 
String theory is one of the most successful theoretical frameworks 
which overcome the difficulty of ultra-violet divergence in quantum
theory of gravity. 
However it has many difficulties in performing practical computations. 
Therefore if we construct a field theory which captures some 
stringy nature and/or includes stringy corrections, then it would play 
a pivotal role in investigating physics in high energy regions 
even near the Planck scale. 

Some of the stringy corrections appear as $\alpha'$ corrections. 
In other words, it often takes the form as higher derivative
corrections i.e. higher order polynomial of momentum.
One of the way to discuss these corrections is deforming the 
Heisenberg uncertainty principle to a generalized uncertainty principle (GUP):
\begin{eqnarray}
\Delta \hat{x} \ge \frac{\hbar}{2\Delta \hat{p}} + \frac{\hbar \beta}{2}\Delta \hat{p},
\end{eqnarray}
where $\beta$ is a deforming parameter.
This parameter has the dimension of
$(length)^2$ and its square root gives the minimal length scale of the
theory under consideration. 
If it is realized in a certain string theory context, $\beta$ would take a value of order the string scale ($\beta \sim \alpha'$). 
This relation comes from various types of studies such as on 
high energy or short distance behavior of 
strings \cite{Gross:1987kz}, \cite{Konishi:1989wk}, gedanken experiment 
of black hole \cite{Maggiore:1993rv}, de Sitter space
\cite{Snyder:1946qz} 
and the symmetry of massless particle \cite{Chagas-Filho:2005at}.

There are several canonical commutation algebra which lead to the GUP. 
Among these algebra we will focus on the algebra:
\begin{eqnarray}
[\hat{x},\hat{p}]=i\hbar (1+\beta \hat{p}^2).
\label{deformed alg}
\end{eqnarray}
This algebra is investigated in \cite{Kempf:1994su}-\cite{Hossenfelder:2005ed} and an attempt to
construct a field theory with minimal length scale is found in
\cite{Kempf:1996ss}. 
In their work they quantize fields using the Bargmann-Fock representation in 1+1 dimensional spacetime.  
This has also been used in cosmology, especially in physics at an early universe 
(see for example, \cite{Ashoorioon:2004vm}-\cite{Ashoorioon:2005ep} and references therein).

In this paper, we investigate the quantization of fields based on the deformed algebra (\ref{deformed alg}) in the canonical formalism in 1+1 dimensions and in the path integral formalism as well. 
In the process of quantization using the latter formalism, we use the
Bjorken-Johnson-Low prescription \cite{Bjorken:1966jh} which states
that T-product of fields must take same value in the two formalism.

This paper is organized as follows. 
In section 2, we first introduce a Hilbert space definitely in terms of the maximally localized states which satisfy minimal uncertainty in position space \cite{Kempf:1994su}.
In section 3, we quantize fields in 1+1 dimensions in the 
canonical formalism, where the maximally localized states are taken to be one particle states. 
In section 4, we construct 1+1 dimensional quantum field theory in the path integral formalism so that it is equivalent to that in the canonical formalism. 
We find that higher dimensional extension is straightforward in the path integral formalism.

\section{Generalized Uncertainty Principle}

We start with the following deformed Heisenberg algebra \cite{Kempf:1994su}:
\begin{eqnarray}
[\hat{x},\hat{p}]=i\hbar (1+\beta \hat{p}^2),
\label{GUP}
\end{eqnarray}
which has a representation in momentum space:
\begin{eqnarray}
\hat{p}&=&p , \nonumber \\
\hat{x}&=&i \hbar (1+\beta p^2) \frac{\partial}{\partial p}.
\end{eqnarray}
In general, states which have minimal uncertainty obey the following equation:
\begin{eqnarray}
\bigl(\hat{x}-\langle \hat{x} \rangle + \frac{\langle \left[ \Delta
						       \hat{x}, \Delta \hat{p}
\right]\rangle}{2(\Delta p)^2}\Delta \hat{p}\bigr)|\psi \rangle &=&0,
\end{eqnarray}
where
\begin{eqnarray}
\Delta \hat{A}&\equiv&\hat{A}-\langle \hat{A} \rangle,
 \nonumber \\
\Delta{A}&\equiv&\sqrt{\langle(\Delta\hat{A})^2 \rangle}. \nonumber
\end{eqnarray}
Among these states, we will focus on the states which are subject to the following conditions: 
\begin{eqnarray}
\Delta x &=& \hbar \sqrt{\beta}, \\
\Delta p &=& \frac{1}{\sqrt{\beta}}, \\
\langle \hat{p} \rangle &=& 0.
\end{eqnarray}

\begin{figure}[h]
\begin{center}
\setlength{\unitlength}{0.240900pt}
\ifx\plotpoint\undefined\newsavebox{\plotpoint}\fi
\sbox{\plotpoint}{\rule[-0.200pt]{0.400pt}{0.400pt}}%
\begin{picture}(1500,900)(0,0)
\font\gnuplot=cmr10 at 10pt
\gnuplot
\sbox{\plotpoint}{\rule[-0.200pt]{0.400pt}{0.400pt}}%
\put(141.0,123.0){\rule[-0.200pt]{4.818pt}{0.400pt}}
\put(121,123){\makebox(0,0)[r]{ 0}}
\put(1419.0,123.0){\rule[-0.200pt]{4.818pt}{0.400pt}}
\put(141.0,246.0){\rule[-0.200pt]{28.818pt}{0.400pt}}
\put(121,246){\makebox(0,0)[r]{ $\sqrt{\beta}$}}
\put(1419.0,246.0){\rule[-0.200pt]{4.818pt}{0.400pt}}
\put(141.0,369.0){\rule[-0.200pt]{4.818pt}{0.400pt}}
\put(121,369){\makebox(0,0)[r]{ 2$\sqrt{\beta}$}}
\put(1419.0,369.0){\rule[-0.200pt]{4.818pt}{0.400pt}}
\put(141.0,491.0){\rule[-0.200pt]{4.818pt}{0.400pt}}
\put(1419.0,491.0){\rule[-0.200pt]{4.818pt}{0.400pt}}
\put(141.0,614.0){\rule[-0.200pt]{4.818pt}{0.400pt}}
\put(121,614){\makebox(0,0)[r]{ 4$\sqrt{\beta}$}}
\put(1419.0,614.0){\rule[-0.200pt]{4.818pt}{0.400pt}}
\put(141.0,737.0){\rule[-0.200pt]{4.818pt}{0.400pt}}
\put(121,737){\makebox(0,0)[r]{ 5$\sqrt{\beta}$}}
\put(1419.0,737.0){\rule[-0.200pt]{4.818pt}{0.400pt}}
\put(141.0,860.0){\rule[-0.200pt]{4.818pt}{0.400pt}}
\put(121,860){\makebox(0,0)[r]{ 6$\sqrt{\beta}$}}
\put(1419.0,860.0){\rule[-0.200pt]{4.818pt}{0.400pt}}
\put(141.0,123.0){\rule[-0.200pt]{0.400pt}{4.818pt}}
\put(141,82){\makebox(0,0){ 0}}
\put(141.0,840.0){\rule[-0.200pt]{0.400pt}{4.818pt}}
\put(271,123.0){\rule[-0.200pt]{0.400pt}{30.318pt}}
\put(265,246){\makebox(0,0){ $\otimes$}}
\put(300,210){\makebox(0,0){ $\nwarrow$}}
\put(590,190){\makebox(0,0){`Maximal localization states'}}
\put(271,82){\makebox(0,0){ $\frac{1}{\sqrt{\beta}}$}}
\put(401.0,123.0){\rule[-0.200pt]{0.400pt}{4.818pt}}
\put(401,82){\makebox(0,0){ $\frac{2}{\sqrt{\beta}}$}}
\put(401.0,840.0){\rule[-0.200pt]{0.400pt}{4.818pt}}
\put(660.0,123.0){\rule[-0.200pt]{0.400pt}{4.818pt}}
\put(660,82){\makebox(0,0){ $\frac{4}{\sqrt{\beta}}$}}
\put(660.0,840.0){\rule[-0.200pt]{0.400pt}{4.818pt}}
\put(920.0,123.0){\rule[-0.200pt]{0.400pt}{4.818pt}}
\put(920,82){\makebox(0,0){ $\frac{6}{\sqrt{\beta}}$}}
\put(920.0,840.0){\rule[-0.200pt]{0.400pt}{4.818pt}}
\put(1179.0,123.0){\rule[-0.200pt]{0.400pt}{4.818pt}}
\put(1179,82){\makebox(0,0){ $\frac{8}{\sqrt{\beta}}$}}
\put(1179.0,840.0){\rule[-0.200pt]{0.400pt}{4.818pt}}
\put(1439.0,123.0){\rule[-0.200pt]{0.400pt}{4.818pt}}
\put(1439,82){\makebox(0,0){ $\frac{10}{\sqrt{\beta}}$}}
\put(1439.0,840.0){\rule[-0.200pt]{0.400pt}{4.818pt}}
\put(141.0,123.0){\rule[-0.200pt]{312.688pt}{0.400pt}}
\put(1439.0,123.0){\rule[-0.200pt]{0.400pt}{177.543pt}}
\put(141.0,860.0){\rule[-0.200pt]{312.688pt}{0.400pt}}
\put(40,491){\makebox(0,0){$\Delta x$}}
\put(790,21){\makebox(0,0){$\Delta p$}}
\put(141.0,123.0){\rule[-0.200pt]{0.400pt}{177.543pt}}
\put(154,737){\usebox{\plotpoint}}
\multiput(154.58,698.52)(0.493,-11.796){23}{\rule{0.119pt}{9.269pt}}
\multiput(153.17,717.76)(13.000,-278.761){2}{\rule{0.400pt}{4.635pt}}
\multiput(167.58,426.45)(0.493,-3.748){23}{\rule{0.119pt}{3.023pt}}
\multiput(166.17,432.73)(13.000,-88.725){2}{\rule{0.400pt}{1.512pt}}
\multiput(180.58,337.96)(0.493,-1.726){23}{\rule{0.119pt}{1.454pt}}
\multiput(179.17,340.98)(13.000,-40.982){2}{\rule{0.400pt}{0.727pt}}
\multiput(193.58,296.74)(0.494,-0.864){25}{\rule{0.119pt}{0.786pt}}
\multiput(192.17,298.37)(14.000,-22.369){2}{\rule{0.400pt}{0.393pt}}
\multiput(207.58,273.80)(0.493,-0.536){23}{\rule{0.119pt}{0.531pt}}
\multiput(206.17,274.90)(13.000,-12.898){2}{\rule{0.400pt}{0.265pt}}
\multiput(220.00,260.93)(0.728,-0.489){15}{\rule{0.678pt}{0.118pt}}
\multiput(220.00,261.17)(11.593,-9.000){2}{\rule{0.339pt}{0.400pt}}
\multiput(233.00,251.94)(1.797,-0.468){5}{\rule{1.400pt}{0.113pt}}
\multiput(233.00,252.17)(10.094,-4.000){2}{\rule{0.700pt}{0.400pt}}
\multiput(246.00,247.95)(2.695,-0.447){3}{\rule{1.833pt}{0.108pt}}
\multiput(246.00,248.17)(9.195,-3.000){2}{\rule{0.917pt}{0.400pt}}
\put(272,245.67){\rule{3.132pt}{0.400pt}}
\multiput(272.00,245.17)(6.500,1.000){2}{\rule{1.566pt}{0.400pt}}
\put(285,246.67){\rule{3.132pt}{0.400pt}}
\multiput(285.00,246.17)(6.500,1.000){2}{\rule{1.566pt}{0.400pt}}
\put(298,248.17){\rule{2.700pt}{0.400pt}}
\multiput(298.00,247.17)(7.396,2.000){2}{\rule{1.350pt}{0.400pt}}
\multiput(311.00,250.61)(2.918,0.447){3}{\rule{1.967pt}{0.108pt}}
\multiput(311.00,249.17)(9.918,3.000){2}{\rule{0.983pt}{0.400pt}}
\multiput(325.00,253.60)(1.797,0.468){5}{\rule{1.400pt}{0.113pt}}
\multiput(325.00,252.17)(10.094,4.000){2}{\rule{0.700pt}{0.400pt}}
\multiput(338.00,257.61)(2.695,0.447){3}{\rule{1.833pt}{0.108pt}}
\multiput(338.00,256.17)(9.195,3.000){2}{\rule{0.917pt}{0.400pt}}
\multiput(351.00,260.60)(1.797,0.468){5}{\rule{1.400pt}{0.113pt}}
\multiput(351.00,259.17)(10.094,4.000){2}{\rule{0.700pt}{0.400pt}}
\multiput(364.00,264.60)(1.797,0.468){5}{\rule{1.400pt}{0.113pt}}
\multiput(364.00,263.17)(10.094,4.000){2}{\rule{0.700pt}{0.400pt}}
\multiput(377.00,268.59)(1.378,0.477){7}{\rule{1.140pt}{0.115pt}}
\multiput(377.00,267.17)(10.634,5.000){2}{\rule{0.570pt}{0.400pt}}
\multiput(390.00,273.60)(1.797,0.468){5}{\rule{1.400pt}{0.113pt}}
\multiput(390.00,272.17)(10.094,4.000){2}{\rule{0.700pt}{0.400pt}}
\multiput(403.00,277.59)(1.378,0.477){7}{\rule{1.140pt}{0.115pt}}
\multiput(403.00,276.17)(10.634,5.000){2}{\rule{0.570pt}{0.400pt}}
\multiput(416.00,282.59)(1.378,0.477){7}{\rule{1.140pt}{0.115pt}}
\multiput(416.00,281.17)(10.634,5.000){2}{\rule{0.570pt}{0.400pt}}
\multiput(429.00,287.59)(1.489,0.477){7}{\rule{1.220pt}{0.115pt}}
\multiput(429.00,286.17)(11.468,5.000){2}{\rule{0.610pt}{0.400pt}}
\multiput(443.00,292.59)(1.378,0.477){7}{\rule{1.140pt}{0.115pt}}
\multiput(443.00,291.17)(10.634,5.000){2}{\rule{0.570pt}{0.400pt}}
\multiput(456.00,297.59)(1.378,0.477){7}{\rule{1.140pt}{0.115pt}}
\multiput(456.00,296.17)(10.634,5.000){2}{\rule{0.570pt}{0.400pt}}
\multiput(469.00,302.59)(1.123,0.482){9}{\rule{0.967pt}{0.116pt}}
\multiput(469.00,301.17)(10.994,6.000){2}{\rule{0.483pt}{0.400pt}}
\multiput(482.00,308.59)(1.378,0.477){7}{\rule{1.140pt}{0.115pt}}
\multiput(482.00,307.17)(10.634,5.000){2}{\rule{0.570pt}{0.400pt}}
\multiput(495.00,313.59)(1.378,0.477){7}{\rule{1.140pt}{0.115pt}}
\multiput(495.00,312.17)(10.634,5.000){2}{\rule{0.570pt}{0.400pt}}
\multiput(508.00,318.59)(1.123,0.482){9}{\rule{0.967pt}{0.116pt}}
\multiput(508.00,317.17)(10.994,6.000){2}{\rule{0.483pt}{0.400pt}}
\multiput(521.00,324.59)(1.378,0.477){7}{\rule{1.140pt}{0.115pt}}
\multiput(521.00,323.17)(10.634,5.000){2}{\rule{0.570pt}{0.400pt}}
\multiput(534.00,329.59)(1.123,0.482){9}{\rule{0.967pt}{0.116pt}}
\multiput(534.00,328.17)(10.994,6.000){2}{\rule{0.483pt}{0.400pt}}
\multiput(547.00,335.59)(1.214,0.482){9}{\rule{1.033pt}{0.116pt}}
\multiput(547.00,334.17)(11.855,6.000){2}{\rule{0.517pt}{0.400pt}}
\multiput(561.00,341.59)(1.378,0.477){7}{\rule{1.140pt}{0.115pt}}
\multiput(561.00,340.17)(10.634,5.000){2}{\rule{0.570pt}{0.400pt}}
\multiput(574.00,346.59)(1.123,0.482){9}{\rule{0.967pt}{0.116pt}}
\multiput(574.00,345.17)(10.994,6.000){2}{\rule{0.483pt}{0.400pt}}
\multiput(587.00,352.59)(1.123,0.482){9}{\rule{0.967pt}{0.116pt}}
\multiput(587.00,351.17)(10.994,6.000){2}{\rule{0.483pt}{0.400pt}}
\multiput(600.00,358.59)(1.378,0.477){7}{\rule{1.140pt}{0.115pt}}
\multiput(600.00,357.17)(10.634,5.000){2}{\rule{0.570pt}{0.400pt}}
\multiput(613.00,363.59)(1.123,0.482){9}{\rule{0.967pt}{0.116pt}}
\multiput(613.00,362.17)(10.994,6.000){2}{\rule{0.483pt}{0.400pt}}
\multiput(626.00,369.59)(1.123,0.482){9}{\rule{0.967pt}{0.116pt}}
\multiput(626.00,368.17)(10.994,6.000){2}{\rule{0.483pt}{0.400pt}}
\multiput(639.00,375.59)(1.123,0.482){9}{\rule{0.967pt}{0.116pt}}
\multiput(639.00,374.17)(10.994,6.000){2}{\rule{0.483pt}{0.400pt}}
\multiput(652.00,381.59)(1.378,0.477){7}{\rule{1.140pt}{0.115pt}}
\multiput(652.00,380.17)(10.634,5.000){2}{\rule{0.570pt}{0.400pt}}
\multiput(665.00,386.59)(1.214,0.482){9}{\rule{1.033pt}{0.116pt}}
\multiput(665.00,385.17)(11.855,6.000){2}{\rule{0.517pt}{0.400pt}}
\multiput(679.00,392.59)(1.123,0.482){9}{\rule{0.967pt}{0.116pt}}
\multiput(679.00,391.17)(10.994,6.000){2}{\rule{0.483pt}{0.400pt}}
\multiput(692.00,398.59)(1.123,0.482){9}{\rule{0.967pt}{0.116pt}}
\multiput(692.00,397.17)(10.994,6.000){2}{\rule{0.483pt}{0.400pt}}
\multiput(705.00,404.59)(1.123,0.482){9}{\rule{0.967pt}{0.116pt}}
\multiput(705.00,403.17)(10.994,6.000){2}{\rule{0.483pt}{0.400pt}}
\multiput(718.00,410.59)(1.123,0.482){9}{\rule{0.967pt}{0.116pt}}
\multiput(718.00,409.17)(10.994,6.000){2}{\rule{0.483pt}{0.400pt}}
\multiput(731.00,416.59)(1.123,0.482){9}{\rule{0.967pt}{0.116pt}}
\multiput(731.00,415.17)(10.994,6.000){2}{\rule{0.483pt}{0.400pt}}
\multiput(744.00,422.59)(1.123,0.482){9}{\rule{0.967pt}{0.116pt}}
\multiput(744.00,421.17)(10.994,6.000){2}{\rule{0.483pt}{0.400pt}}
\multiput(757.00,428.59)(1.378,0.477){7}{\rule{1.140pt}{0.115pt}}
\multiput(757.00,427.17)(10.634,5.000){2}{\rule{0.570pt}{0.400pt}}
\multiput(770.00,433.59)(1.123,0.482){9}{\rule{0.967pt}{0.116pt}}
\multiput(770.00,432.17)(10.994,6.000){2}{\rule{0.483pt}{0.400pt}}
\multiput(783.00,439.59)(1.214,0.482){9}{\rule{1.033pt}{0.116pt}}
\multiput(783.00,438.17)(11.855,6.000){2}{\rule{0.517pt}{0.400pt}}
\multiput(797.00,445.59)(1.123,0.482){9}{\rule{0.967pt}{0.116pt}}
\multiput(797.00,444.17)(10.994,6.000){2}{\rule{0.483pt}{0.400pt}}
\multiput(810.00,451.59)(1.123,0.482){9}{\rule{0.967pt}{0.116pt}}
\multiput(810.00,450.17)(10.994,6.000){2}{\rule{0.483pt}{0.400pt}}
\multiput(823.00,457.59)(1.123,0.482){9}{\rule{0.967pt}{0.116pt}}
\multiput(823.00,456.17)(10.994,6.000){2}{\rule{0.483pt}{0.400pt}}
\multiput(836.00,463.59)(1.123,0.482){9}{\rule{0.967pt}{0.116pt}}
\multiput(836.00,462.17)(10.994,6.000){2}{\rule{0.483pt}{0.400pt}}
\multiput(849.00,469.59)(1.123,0.482){9}{\rule{0.967pt}{0.116pt}}
\multiput(849.00,468.17)(10.994,6.000){2}{\rule{0.483pt}{0.400pt}}
\multiput(862.00,475.59)(1.123,0.482){9}{\rule{0.967pt}{0.116pt}}
\multiput(862.00,474.17)(10.994,6.000){2}{\rule{0.483pt}{0.400pt}}
\multiput(875.00,481.59)(1.123,0.482){9}{\rule{0.967pt}{0.116pt}}
\multiput(875.00,480.17)(10.994,6.000){2}{\rule{0.483pt}{0.400pt}}
\multiput(888.00,487.59)(1.123,0.482){9}{\rule{0.967pt}{0.116pt}}
\multiput(888.00,486.17)(10.994,6.000){2}{\rule{0.483pt}{0.400pt}}
\multiput(901.00,493.59)(1.214,0.482){9}{\rule{1.033pt}{0.116pt}}
\multiput(901.00,492.17)(11.855,6.000){2}{\rule{0.517pt}{0.400pt}}
\multiput(915.00,499.59)(1.123,0.482){9}{\rule{0.967pt}{0.116pt}}
\multiput(915.00,498.17)(10.994,6.000){2}{\rule{0.483pt}{0.400pt}}
\multiput(928.00,505.59)(1.123,0.482){9}{\rule{0.967pt}{0.116pt}}
\multiput(928.00,504.17)(10.994,6.000){2}{\rule{0.483pt}{0.400pt}}
\multiput(941.00,511.59)(1.123,0.482){9}{\rule{0.967pt}{0.116pt}}
\multiput(941.00,510.17)(10.994,6.000){2}{\rule{0.483pt}{0.400pt}}
\multiput(954.00,517.59)(1.123,0.482){9}{\rule{0.967pt}{0.116pt}}
\multiput(954.00,516.17)(10.994,6.000){2}{\rule{0.483pt}{0.400pt}}
\multiput(967.00,523.59)(0.950,0.485){11}{\rule{0.843pt}{0.117pt}}
\multiput(967.00,522.17)(11.251,7.000){2}{\rule{0.421pt}{0.400pt}}
\multiput(980.00,530.59)(1.123,0.482){9}{\rule{0.967pt}{0.116pt}}
\multiput(980.00,529.17)(10.994,6.000){2}{\rule{0.483pt}{0.400pt}}
\multiput(993.00,536.59)(1.123,0.482){9}{\rule{0.967pt}{0.116pt}}
\multiput(993.00,535.17)(10.994,6.000){2}{\rule{0.483pt}{0.400pt}}
\multiput(1006.00,542.59)(1.123,0.482){9}{\rule{0.967pt}{0.116pt}}
\multiput(1006.00,541.17)(10.994,6.000){2}{\rule{0.483pt}{0.400pt}}
\multiput(1019.00,548.59)(1.214,0.482){9}{\rule{1.033pt}{0.116pt}}
\multiput(1019.00,547.17)(11.855,6.000){2}{\rule{0.517pt}{0.400pt}}
\multiput(1033.00,554.59)(1.123,0.482){9}{\rule{0.967pt}{0.116pt}}
\multiput(1033.00,553.17)(10.994,6.000){2}{\rule{0.483pt}{0.400pt}}
\multiput(1046.00,560.59)(1.123,0.482){9}{\rule{0.967pt}{0.116pt}}
\multiput(1046.00,559.17)(10.994,6.000){2}{\rule{0.483pt}{0.400pt}}
\multiput(1059.00,566.59)(1.123,0.482){9}{\rule{0.967pt}{0.116pt}}
\multiput(1059.00,565.17)(10.994,6.000){2}{\rule{0.483pt}{0.400pt}}
\multiput(1072.00,572.59)(1.123,0.482){9}{\rule{0.967pt}{0.116pt}}
\multiput(1072.00,571.17)(10.994,6.000){2}{\rule{0.483pt}{0.400pt}}
\multiput(1085.00,578.59)(1.123,0.482){9}{\rule{0.967pt}{0.116pt}}
\multiput(1085.00,577.17)(10.994,6.000){2}{\rule{0.483pt}{0.400pt}}
\multiput(1098.00,584.59)(1.123,0.482){9}{\rule{0.967pt}{0.116pt}}
\multiput(1098.00,583.17)(10.994,6.000){2}{\rule{0.483pt}{0.400pt}}
\multiput(1111.00,590.59)(1.123,0.482){9}{\rule{0.967pt}{0.116pt}}
\multiput(1111.00,589.17)(10.994,6.000){2}{\rule{0.483pt}{0.400pt}}
\multiput(1124.00,596.59)(1.123,0.482){9}{\rule{0.967pt}{0.116pt}}
\multiput(1124.00,595.17)(10.994,6.000){2}{\rule{0.483pt}{0.400pt}}
\multiput(1137.00,602.59)(1.026,0.485){11}{\rule{0.900pt}{0.117pt}}
\multiput(1137.00,601.17)(12.132,7.000){2}{\rule{0.450pt}{0.400pt}}
\multiput(1151.00,609.59)(1.123,0.482){9}{\rule{0.967pt}{0.116pt}}
\multiput(1151.00,608.17)(10.994,6.000){2}{\rule{0.483pt}{0.400pt}}
\multiput(1164.00,615.59)(1.123,0.482){9}{\rule{0.967pt}{0.116pt}}
\multiput(1164.00,614.17)(10.994,6.000){2}{\rule{0.483pt}{0.400pt}}
\multiput(1177.00,621.59)(1.123,0.482){9}{\rule{0.967pt}{0.116pt}}
\multiput(1177.00,620.17)(10.994,6.000){2}{\rule{0.483pt}{0.400pt}}
\multiput(1190.00,627.59)(1.123,0.482){9}{\rule{0.967pt}{0.116pt}}
\multiput(1190.00,626.17)(10.994,6.000){2}{\rule{0.483pt}{0.400pt}}
\multiput(1203.00,633.59)(1.123,0.482){9}{\rule{0.967pt}{0.116pt}}
\multiput(1203.00,632.17)(10.994,6.000){2}{\rule{0.483pt}{0.400pt}}
\multiput(1216.00,639.59)(1.123,0.482){9}{\rule{0.967pt}{0.116pt}}
\multiput(1216.00,638.17)(10.994,6.000){2}{\rule{0.483pt}{0.400pt}}
\multiput(1229.00,645.59)(1.123,0.482){9}{\rule{0.967pt}{0.116pt}}
\multiput(1229.00,644.17)(10.994,6.000){2}{\rule{0.483pt}{0.400pt}}
\multiput(1242.00,651.59)(1.123,0.482){9}{\rule{0.967pt}{0.116pt}}
\multiput(1242.00,650.17)(10.994,6.000){2}{\rule{0.483pt}{0.400pt}}
\multiput(1255.00,657.59)(1.026,0.485){11}{\rule{0.900pt}{0.117pt}}
\multiput(1255.00,656.17)(12.132,7.000){2}{\rule{0.450pt}{0.400pt}}
\multiput(1269.00,664.59)(1.123,0.482){9}{\rule{0.967pt}{0.116pt}}
\multiput(1269.00,663.17)(10.994,6.000){2}{\rule{0.483pt}{0.400pt}}
\multiput(1282.00,670.59)(1.123,0.482){9}{\rule{0.967pt}{0.116pt}}
\multiput(1282.00,669.17)(10.994,6.000){2}{\rule{0.483pt}{0.400pt}}
\multiput(1295.00,676.59)(1.123,0.482){9}{\rule{0.967pt}{0.116pt}}
\multiput(1295.00,675.17)(10.994,6.000){2}{\rule{0.483pt}{0.400pt}}
\multiput(1308.00,682.59)(1.123,0.482){9}{\rule{0.967pt}{0.116pt}}
\multiput(1308.00,681.17)(10.994,6.000){2}{\rule{0.483pt}{0.400pt}}
\multiput(1321.00,688.59)(1.123,0.482){9}{\rule{0.967pt}{0.116pt}}
\multiput(1321.00,687.17)(10.994,6.000){2}{\rule{0.483pt}{0.400pt}}
\multiput(1334.00,694.59)(1.123,0.482){9}{\rule{0.967pt}{0.116pt}}
\multiput(1334.00,693.17)(10.994,6.000){2}{\rule{0.483pt}{0.400pt}}
\multiput(1347.00,700.59)(1.123,0.482){9}{\rule{0.967pt}{0.116pt}}
\multiput(1347.00,699.17)(10.994,6.000){2}{\rule{0.483pt}{0.400pt}}
\multiput(1360.00,706.59)(0.950,0.485){11}{\rule{0.843pt}{0.117pt}}
\multiput(1360.00,705.17)(11.251,7.000){2}{\rule{0.421pt}{0.400pt}}
\multiput(1373.00,713.59)(1.214,0.482){9}{\rule{1.033pt}{0.116pt}}
\multiput(1373.00,712.17)(11.855,6.000){2}{\rule{0.517pt}{0.400pt}}
\multiput(1387.00,719.59)(1.123,0.482){9}{\rule{0.967pt}{0.116pt}}
\multiput(1387.00,718.17)(10.994,6.000){2}{\rule{0.483pt}{0.400pt}}
\multiput(1400.00,725.59)(1.123,0.482){9}{\rule{0.967pt}{0.116pt}}
\multiput(1400.00,724.17)(10.994,6.000){2}{\rule{0.483pt}{0.400pt}}
\multiput(1413.00,731.59)(1.123,0.482){9}{\rule{0.967pt}{0.116pt}}
\multiput(1413.00,730.17)(10.994,6.000){2}{\rule{0.483pt}{0.400pt}}
\multiput(1426.00,737.59)(1.123,0.482){9}{\rule{0.967pt}{0.116pt}}
\multiput(1426.00,736.17)(10.994,6.000){2}{\rule{0.483pt}{0.400pt}}
\put(259.0,246.0){\rule[-0.200pt]{3.132pt}{0.400pt}}
\end{picture}
\end{center}
\caption{{Generalized uncertainty relation of $x-p$}}
\label{Fig:GUP}
\end{figure}
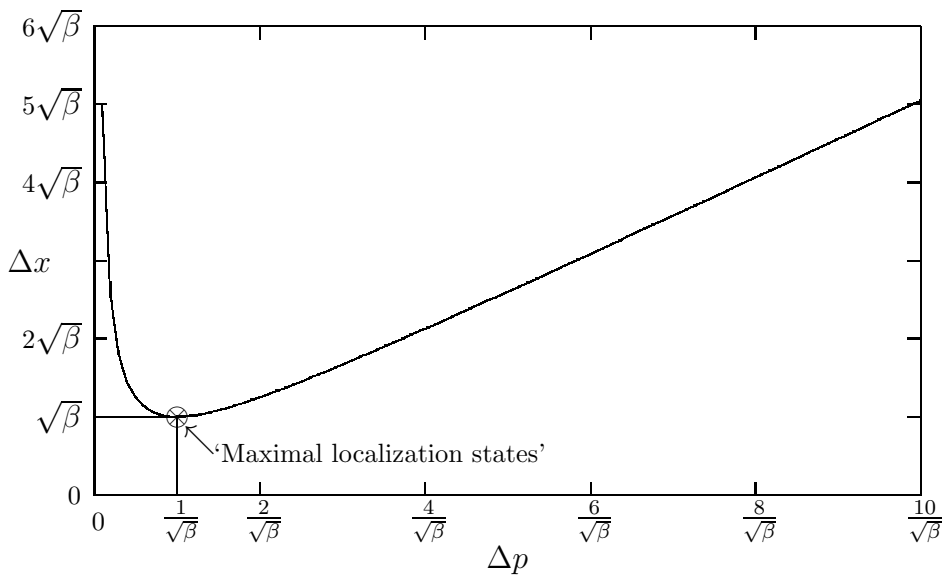
As is obvious in Figure \ref{Fig:GUP}, 
these states realize the minimal value of $\Delta x$ 
and are called as `maximal localization states' \cite{Kempf:1994su}.
One can easily find the wavefunction of maximal localization states in momentum space as
\begin{eqnarray}
\psi_{\xi}(p)=\frac{N_1}{\sqrt{2\pi \hbar}}\frac{1}{\sqrt{1+\beta p^2}}\exp\left(\frac{\xi}{i\hbar\sqrt{\beta}}
\tan^{-1}(\sqrt{\beta}p)\right), \label{wf}
\end{eqnarray}
where $\xi \equiv \langle \hat{x} \rangle$ and $N_1$ is a normalization 
factor. 
These maximal localization states are identified as `one particle state' in
the theory based on GUP \cite{Kempf:1994su}.
We will use these states as basic ingredients in constructing fields in the later section. 
Note that these states are not eigenstates of the position operator $\hat{x}$.

We here introduce a first quantized Hilbert space as the space equipped with the complete basis $\{|p\rangle\}$ which has the completeness:
\begin{eqnarray}
\vmi{ 1} = \int {\cal D}p \sp |p\rangle \langle p |.
\end{eqnarray}
We determine the measure ${\cal D}p$ so that the operator 
$\hat{x}= i\hbar (1+\beta p^2) \frac{\partial}{\partial p}$ is hermitian
($\langle \chi|\hat{x}\psi\rangle=\langle \hat{x}\chi |\psi\rangle$):
\begin{eqnarray}
{\cal D}p \equiv N_2 \frac{dp}{1+\beta p^2},\label{pmeasure}
\end{eqnarray}
where $N_2$ is a normalization factor independent of $p$.
We set $N_2=1$ by rescaling the basis as $\sqrt{N_2}|p\rangle \to |p\rangle$.
This measure leads to the orthogonal property:
\begin{eqnarray}
\langle p| p' \rangle = (1+\beta p^2)\delta(p-p').
\end{eqnarray}
We write maximal localization states as $|\psi_\xi \rangle$, then we can 
rewrite the wavefunction of one particle state (\ref{wf}) in the
bracket notation: 
\begin{eqnarray}
\psi_{\xi}(p)=\langle p|\psi_\xi \rangle.
\end{eqnarray}
In this Hilbert space, the inner product of one particle states is 
\begin{eqnarray}
\langle \psi_{\xi'} | \psi_\xi \rangle &=&\int^{\infty}_{-\infty} 
       \frac{dp}{1+\beta p^2}
       \psi_{\xi'}^{*}(p)\psi_{\xi}(p) \nonumber \\
 &=&\frac{|N_1|^2}{2\pi\hbar}\frac{1}{2\sqrt{\beta}\left(\frac{\xi-\xi'}{2\hbar \sqrt{\beta}}-(\frac{\xi-\xi'}{2\hbar \sqrt{\beta}})^3\right)}\sin\left(\frac{\xi-\xi'}{2\hbar \sqrt{\beta}}\pi\right), 
\label{IP}
\end{eqnarray}
which shows they are not completely orthogonal with each other. 
This is because these states have $\Delta x_{\mbox{min}} \ne 0$.   

Furthermore, we find a completeness of basis $\{|\psi_\xi \rangle\}$ on this Hilbert space as
\begin{eqnarray}
\vmi{ 1} &=& \int \frac{d\xi}{|N_1|^2}(1+\beta
 \hat{p}^2)|\psi_\xi\rangle\langle \psi_\xi| \nonumber \\
 &=& \int \frac{d\xi}{|N_1|^2}|\psi_\xi\rangle\langle \psi_\xi|
(1+\beta \hat{p}^2).
\end{eqnarray}
We set $N_1=1$ by rescaling the basis as 
$(N_1)^{-1}|\psi_\xi \rangle \to |\psi_\xi\rangle$.
        
Here we have two bases $\{|p\rangle\}, \{|\psi_\xi \rangle\}$. 
Thus we can perform a `Fourier transformation' which interchanges
the two representations.
We write the $p$-space wavefunction and $\xi$-space wavefunction 
for a state $|\phi\rangle$ in this Hilbert space as follows,
\footnote{
We use the same character $\phi$ for $\xi$-space and $p$-space
wavefunctions, 
because we can easily recognize on which space a field lives 
by noting the argument of the function. 
}
\begin{eqnarray}
 \phi(p)&\equiv& \langle p|\phi\rangle, \nonumber \\
 \phi(\xi) &\equiv& \langle \psi_\xi|\phi\rangle.
\end{eqnarray}
The Fourier transformation in this Hilbert space is 
\begin{eqnarray}
\label{FT1}
\phi(\xi)&=& \int^{\infty}_{-\infty}\frac{dp}{1+\beta p^2}
            \psi_{\xi}^{*}(p)\phi(p), \\
\phi(p)&=&\int^{\infty}_{-\infty} d\xi (1+\beta
 p^2)\psi_{\xi}(p)\phi(\xi).
\label{FT2}
\end{eqnarray}

Here there are statements that are worth while mentioning. 
The transformation (\ref{FT2}) is an inverse transformation to the
one (\ref{FT1}) for arbitrary square integrable functions like as in the case of usual Fourier transformation.
However the transformation (\ref{FT1}) is not an inverse transformation to the one (\ref{FT2}) for arbitrary square integrable functions.
This can be seen by inserting (\ref{FT2}) into (\ref{FT1}), then one finds
\begin{eqnarray}
\int^{\infty}_{-\infty}\frac{dp}{1+\beta p^2}
            \psi_{\xi}^{*}(p)\phi(p)&=&\int^{\infty}_{-\infty}\frac{dp}{1+\beta p^2}
            \psi_{\xi}^{*}(p)\int^{\infty}_{-\infty} d\xi' (1+\beta
 p^2)\psi_{\xi'}(p)\phi(\xi') \nonumber \\
&=&\int^{\infty}_{-\infty}\frac{d\xi'}{\pi}\frac{\hbar}{\xi-\xi'}\sin\left(\frac{(\xi-\xi')\pi}{2\hbar\sqrt{\beta}}\right)\phi(\xi') .
\end{eqnarray}
This is not equal to the original function for arbitrary function $\phi(\xi')$.

Here we must restrict the function $\phi(\xi')$ to linear
combination of $\psi^*_{\xi'}(p)$'s as 
$\phi(\xi')=\int \frac{dp}{1+\beta p^2} a(p)\psi_{\xi'}^*(p)$.
For these functions the transformation (\ref{FT1}) is truly the inverse transformation to (\ref{FT2}):
\begin{eqnarray}
\int^{\infty}_{-\infty}\frac{dp}{1+\beta p^2}
            \psi_{\xi}^{*}(p)\phi(p)&=&\int^{\infty}_{-\infty}\frac{dp}{1+\beta p^2}
            \psi_{\xi}^{*}(p)\int^{\infty}_{-\infty} d\xi' (1+\beta
 p^2)\psi_{\xi'}(p)\phi(\xi') \nonumber \\
&=& \int^{\infty}_{-\infty}\frac{dp'dpd\xi'}{1+\beta p'^2} a(p')
            \psi_{\xi}^{*}(p) \psi_{\xi'}(p)\psi^*_{\xi'}(p') \nonumber \\
&=& \int^{\infty}_{-\infty}\frac{dp'}{1+\beta p'^2} a(p')\psi_{\xi}^*(p') \nonumber \\
&=& \phi(\xi).
\end{eqnarray}
This restriction is nothing but the condition that the state $|\phi\rangle$ is included in our Hilbert space. 
Thus the Fourier transformations (\ref{FT1}) and (\ref{FT2}) are well-defined in this Hilbert space.  

\begin{figure}[h]
\begin{center}
\resizebox{10cm}{10cm}{\includegraphics{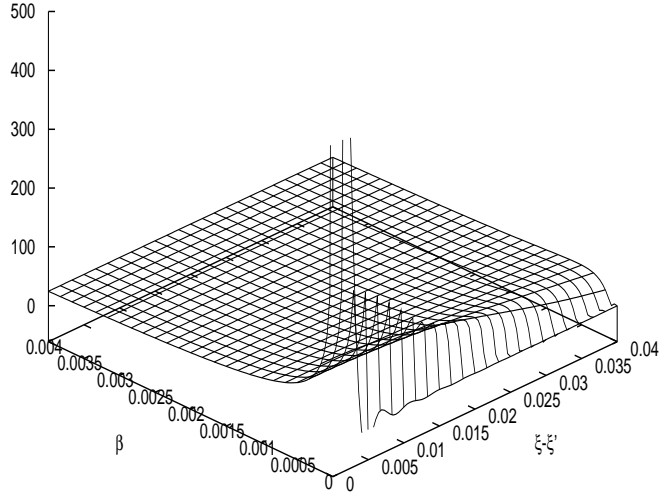}}
\end{center}
\caption{{Shape of $\langle\psi_\xi | \psi_{\xi'}\rangle$} in units $\hbar=1$ and $N_1=1$.}
\label{fig:deltax}
\end{figure}

It should be noted that the wavefunction (\ref{wf}) reduces to the plane wave in the limit $\beta \to 0$:
\begin{equation}
\psi_\xi(p) \to \frac{1}{\sqrt{2\pi \hbar}}e^{-ip\xi/\hbar} .
\end{equation}
This is because we have set the normalization constant as $N_1 = 1$ 
by rescaling $|\psi_\xi \rangle$. 
In \cite{Kempf:1994su} the normalization constant is fixed to $N_1=2\sqrt{\hbar\sqrt{\beta}}$ so that the norm of the wavefunction is equal to one.
In this case, the wavefunction vanishes in the limit
because of the extra dependence on $\beta$ in the normalization factor.
In other words, we regard the wavefunction (\ref{wf}) as if it is non-normalizable, because we have taken it as a counterpart of the plane wave in ordinary quantum mechanics.
The plane wave is non-normalizable.
Once we turn on $\beta$, a nonlocality emerges 
as one can easily see in Figure 2 which shows the behavior of the
inner product (\ref{IP}).

\section{1+1 dimensional quantum field theory on GUP}

In this section we construct a quantum field theory on GUP in 1+1 dimensional spacetime in the canonical formalism. 

First of all, we define the classical field such that it solves the following
Klein-Gordon equation:
\begin{eqnarray}
\label{KG}
0=\left(\left(\hbar\frac{\partial}{\partial t}\right)^2+F(p)+m^2\right)\Phi(p,t).
\end{eqnarray}
In this equation we leave the function of momentum $F(p)$ indefinite,
where $F(p)$ is an arbitrary even function whose explicit form does not have any influence to the following arguments.
This function determines the dispersion relation and depends on what kind of theory we want. 
 
Quantization of a classical field is to provide the set of operator field and Hamiltonian which lead the same Klein-Gordon equation by the Heisenberg equation. 
We define a field $\hat{\Phi}(p,t)$:
\begin{eqnarray}
\label{field}
\hat{\Phi}(p,t)\equiv\frac{\hbar}{\sqrt{E(p)}}\hat{\phi}(p)
\exp(\frac{t}{i \hbar} E(p))
+
\frac{\hbar}{\sqrt{E(-p)}}\hat{\phi}^{\dagger}(-p)
\exp(-\frac{t}{i \hbar} E(-p)) ,
\end{eqnarray}
where $E(p) \equiv \sqrt{F(p)+m^2}$.
This field satisfies the Klein-Gordon equation (\ref{KG}).
Its Fourier pair $\hat{\Phi}(\xi,t)$ is written in terms of  our `one particle states':
\begin{eqnarray}
\hat{\Phi}(\xi,t)&\equiv&\int^{\infty}_{-\infty}\frac{dp}{1+\beta p^2}\{
                         \psi_{\xi}^{*}(p)\hat{\phi}(p)\frac{\hbar}{\sqrt{2E(p)}}
                         \exp(\frac{1}{i\hbar}tE(p)) \nonumber \\
&&+\psi_{\xi}(p)
                         \hat{\phi}^{\dagger}(p)\frac{\hbar}{\sqrt{2E(p)}}
                         \exp(-\frac{1}{i\hbar}tE(p))\} .
\end{eqnarray}
In the above expression, the operators
$\hat{\phi}(p),\hat{\phi}^{\dagger}(p)$ are annihilation and creation
operators for one particle state with momentum $p$.
They have the following commutation relation:
\begin{eqnarray}
\label{COM}
\left[\hat{\phi}(p),\hat{\phi}^{\dagger}(p')\right]=(1+\beta p^2)\delta(p-p').
\end{eqnarray}
At first sight it is unclear whether the generalized canonical commutation relations is the same form as those of ordinary quantum field theory. 
Thus the commutation relation might have an arbitrary power of $(1+\beta p^2)$. 
The factor $(1+\beta p^2)$ in front of the delta function is determined 
by the form of the measure ${\cal{D}}p$ which is the consequence of the
hermiticity of $\hat{x}$ in the first quantized Hilbert space. 

We construct a natural Hamiltonian in terms of the Fock space picture: 
\begin{eqnarray}
{\cal H} \equiv \int \frac{dp}{1+\beta p^2}E(p)\hat{\phi}^{\dagger}(p)\hat{\phi}(p).
\end{eqnarray}
This Hamiltonian produces the Heisenberg equation:
\begin{eqnarray}
\frac{1}{i\hbar}\left[\hat{\Phi}(p,t),{\cal
		 H}\right]=\frac{\partial}{\partial t}\hat{\Phi}(p,t),
\end{eqnarray}
which is consistent with the Klein-Gordon equation (\ref{KG}).
We can define conjugate momentum 
$\hat{\Pi}(p,t) \equiv \frac{\partial}{\partial t}\hat{\Phi}(p,t)$.
We can write the Hamiltonian
\begin{eqnarray}
 {\cal H}&=& \int^{\infty}_{-\infty}\frac{dp}{1+\beta p^2}
 \frac{1}{2}\left[\hat{\Pi}(-p,t)\hat{\Pi}(p,t) 
 +\left(\frac{E(p)}{\hbar}\right)^2\hat{\Phi}(-p,t)\hat{\Phi}(p,t)\right]
\nonumber \\
&&+\mbox{const.}.
\label{HHamil0}
\end{eqnarray}

We compute the following commutation relation by using the commutation relation (\ref{COM}):
\begin{eqnarray}
\left[\hat{\Phi}(p,t),\hat{\Pi}(p',t)\right]&=&i\hbar(1+\beta p^2)
\delta(p+p'), \\
\left[\hat{\Phi}(\xi,t),\hat{\Pi}(\xi' ,t)\right]&=&i\hbar
\langle\psi_\xi|\psi_{\xi'}\rangle.
\end{eqnarray}
As we can see in Figure \ref{fig:deltax}, this commutation relation in $\xi$ space representation does not provide a delta function. 
There is a nonlocality in the quantum field theory like as in the first quantized system which is constructed in the previous section.

\section{Higher dimensional algebra}

In this section we extend the previous argument of the quantization of fields to the case in higher dimensional spacetime. 
In the process of the extension in a naive way, one would be faced with the  difficulty that we can not identify the maximal localization states. 
Therefore we will not use the canonical formalism, instead we use the path integral formalism in which it turns out that we can easily extend to higher dimensional spacetime.

To see the difficulty clearly, we start with higher dimensional GUT algebra.
Jacobi identity determines the full algebra:
\begin{eqnarray}
 \left[\hat{x}^i,\hat{p}_j\right]&=&i\hbar (1+\beta \mbox{\boldmath
  $\hat{p}$}^2)\delta^i_{\sp j}, \label{GUP2}\\
 \left[\hat{x}^i,\hat{x}^j\right]&=&-2i\hbar \beta(1+\beta \mbox{\boldmath $\hat{p}$}^2)\hat{L}^{ij}. 
\label{GUP3}
\end{eqnarray}
Here $\hat{L}^{ij}$ are angular momentum operators 
$\hat{L}^{ij}\equiv \frac{1}{1+\beta \mbox{\boldmath $\hat{p}$}^2}(\hat{x}^i\hat{p}^j-\hat{x}^j\hat{p}^i)$ 
and $i$ runs from 1 to $d$ which is the number of spatial coordinates
and $\mbox{\boldmath$\hat{p}$}^2 \equiv \sum_{i=1}^d (\hat{p}_i)^2$ \cite{Kempf:1994su}.

Let us try to define maximal localization states like 
as we did in 1+1 dimensional case.
Such states would satisfy the equation:
\begin{eqnarray}
\left( i\hbar(1+\beta \mbox{\boldmath $p$}^2)\frac{\partial}{\partial p_i}-\langle
 \hat{x}^i \rangle+ 
 i\hbar\frac{1+\beta (\Delta p_i)^2+\beta \langle p_i \rangle^2}{2\Delta
 p^2}(p_i-\langle p_i\rangle)\right)\psi(p) 
=0. 
\end{eqnarray} 
However these $d$ relations can not coexist because of the noncommutativity of the differential operators acting on the $\psi(p)$. 
In general, the wavefunction which obeys the above equations corresponds to $s$-wave which has zero angular momentum. 
We can not produce states with non zero angular momentum 
from these states only. 
Thus we find it difficult to construct a quantum field theory in the same manner as in the 1+1 dimensional case.

A different way of extension of the GUP algebra (\ref{GUP}) is simply to take a direct product of it:
\begin{eqnarray}
\left[\hat{x}^i,\hat{p}_j\right]=i\hbar (1+\beta (p_i)^2)\delta^i_{\sp j} .
\end{eqnarray}
In this case we can construct a quantum field theory by taking a direct product of fields \cite{Kempf:1996ss}.
However this extension is not a Lorentz invariant deformation.
In particular, the rotational invariance is completely broken. 
Thus here we take the algebra (\ref{GUP2})-(\ref{GUP3}) and investigate another way to construct a quantum field theory based on this algebra. 

To this end, we reconstruct a quantum field theory on GUP in 1+1 dimensions in a  different manner; the path integral formalism in order to be applicable to higher dimensional spacetime. 
In fact, as we shall see later, it is not difficult to extend to higher dimensions in the path integral formalism. 
In a way to construct a field theory in the path integral formalism, 
we adopt a guiding principle that it produces the same result 
as that in the canonical formalism. 
The equivalence of the canonical formalism and the path integral one is
guaranteed by the Bjorken-Johnson-Low (BJL) prescription
\cite{Bjorken:1966jh}. At first we introduce deformed factors of the form 
$(1+\beta p^2)^{\alpha}$ where $\alpha$ denotes free parameters 
in the definitions of Hamiltonian, conjugate momentum fields, Lagrangian and so on. 
This is because in generalized canonical commutation relations it is 
unclear whether the definition of these phase space structure are the
same as those of ordinary quantum field theory. These free parameters
should be determined so that path integral formalism leads the same
result as that of canonical formalism in 1+1 dimensions.


We define path integral measure and Lagrangian by introducing two parameters $l$ and $m$ as follows,
\begin{eqnarray}
1&=& \int {\cal D}\Phi(p,t)\exp\left(-\frac{1}{2}\int \frac{dtdp}{(1+\beta
 p^2)^{l}}\Phi(p,t)\Phi(-p,t)\right), \label{measure}\\
{\cal L}&=& -\frac{1}{2}\int^{\infty}_{-\infty}dp(1+\beta
    p^2)^{m}\Phi(-p,t)\left[\partial_t^2+\left(\frac{E(p)}{\hbar}\right)^2\right]
\Phi(p,t).\label{lag1}
\end{eqnarray}
Then the action $S\equiv\int dt {\cal L}$ is written as,
\begin{eqnarray}
S&=& -\frac{1}{2}\int^{\infty}_{-\infty}dpdp'dqdq'(1+\beta p^2)^{m}
\delta(p+p')\delta(q+q')\nonumber \\
&&\times  \Phi(p,q)\left[-\left(\frac{q}{\hbar}\right)^2+\left(\frac{E(p)}{\hbar}\right)^2\right]\Phi(p',q').
\label{lag2}
\end{eqnarray}
According to the assumptions (\ref{measure}) and (\ref{lag2}) we can compute the $T^*$-product:  
\begin{eqnarray}
\langle T^*\hat{\Phi}(p,q)\hat{\Phi}(p',q')\rangle =\frac{\hbar^3}{i} (1+\beta p^2)^{-m-l}
  \delta(p+p')\delta(q+q')\frac{1}{-q^2+E(p)^2-i\epsilon}. 
\label{LT}
\end{eqnarray}
Using the Bjorken-Johnson-Low prescription, from behavior 
of 
\begin{equation}
\lim_{q \to \infty}\langle T^*\hat{\Phi}(p,q)\hat{\Phi}(p',q')\rangle 
\quad {\mbox{and}} \quad
\lim_{q \to \infty}q\langle T^*\hat{\Phi}(p,q)\hat{\Phi}(p',q')\rangle ,
\end{equation}
we find 
\begin{eqnarray}
T^*\hat{\Phi}(p,q)\hat{\Phi}(p',q')&=&T\hat{\Phi}(p,q)\hat{\Phi}(p',q'), \\
\left[\hat{\Phi}(p,t) , \hat{\Phi}(p',t)\right]&=& 0.
\end{eqnarray}
In the same manner, from behavior of 
\begin{equation}
\lim_{q \to \infty}q^2\langle T^*\hat{\Phi}(p,q)\hat{\Phi}(p',q')\rangle ,
\end{equation}
we also find
\begin{eqnarray}
\left[ \hat{\Phi}(p,t), \dot{\hat{\Phi}}(p',t)\right]=i\hbar
(1+\beta p^2)^{-m-l}\delta(p+p'). \label{Lcomm1}
\end{eqnarray}
If we define conjugate momentum $\Pi(p,t)$ in Lagrangian formalism by using a free parameter $r$ as
\begin{eqnarray}
\Pi(p,t) \equiv (1+\beta p^2)^r\frac{\delta S}{\delta \dot{\Phi}(-p,t)},
\label{Lpi}
\end{eqnarray}
the commutation relation (\ref{Lcomm1}) becomes the following,
\begin{eqnarray}
\left[\hat{\Phi}(p,t),\hat{\Pi}(p',t)\right]=i\hbar(1+\beta p^2)^{-2m-l-r}
\delta(p+p'). \label{Lcomm2}
\end{eqnarray}
The right hand side of the above equation is a symplectic form.
Because the Legendre transformation from Lagrangian to Hamiltonian is 
\begin{eqnarray}
\frac{1}{i\hbar}\int dt {\cal H} \equiv (\mbox{symplectic
 form})^{-1}\sum_i \Pi_i d\Phi_i - \frac{1}{i\hbar}\int dt {\cal L},
\end{eqnarray}
we can obtain a Hamiltonian:
\begin{eqnarray}
{\cal H}&=&\int^{\infty}_{-\infty}dp(1+\beta
    p^2)^{m+l}\left[(1-\frac{1}{2}(1+\beta
    p^2)^{-l-2m-2r})\hat{\Pi}(-p,t)\hat{\Pi}(p,t) 
    \right.
     \nonumber \\
 &&\left.
 +\frac{1}{2}(1+\beta p^2)^{-l}\left(\frac{E(p)}{\hbar}\right)^2\hat{\Phi}(-p,t)\hat{\Phi}(p,t)\right].
\end{eqnarray}
The condition that the coefficient of $\Pi (-p,t)\Pi(p,t)$ must be homogeneous fixes the free parameter $r$:
\begin{eqnarray}
r=-\frac{1}{2}l-m. \label{rfix}
\end{eqnarray}
Then the Hamiltonian becomes 
\begin{eqnarray}
{\cal H}&=&\int^{\infty}_{-\infty}dp(1+\beta
    p^2)^{m+l}\frac{1}{2}\left[\hat{\Pi}(-p,t)\hat{\Pi}(p,t) 
    \right. 
    \nonumber \\
 &&\left.+(1+\beta p^2)^{-l}\left(\frac{E(p)}{\hbar}\right)^2\hat{\Phi}(-p,t)\hat{\Phi}(p,t)\right].\label{LHamil}
\end{eqnarray}
By comparing this with (\ref{HHamil0}), we obtain the parameters $m$ and
$l$ as the form:
\begin{eqnarray}
l=0, \label{lfix} \\
m=-1. \label{mfix}
\end{eqnarray}
 
Now it is apparent the two formalism provide consistent structure of 
the deformed quantum field theory.
Let us summarize the result.
The Hamiltonian is 
\begin{eqnarray}
 {\cal H}&=& \int^{\infty}_{-\infty}\frac{dp}{1+\beta
    p^2}\frac{1}{2}\left[\hat{\Pi}(-p,t)\hat{\Pi}(p,t) 
    \right. \nonumber \\
 &&\left.
 +\left(\frac{E(p)}{\hbar}\right)^2\hat{\Phi}(-p,t)\hat{\Phi}(p,t)\right]+
  \mbox{const.}  \label{finalHam}\\
&=& \int^{\infty}_{-\infty}\frac{dp}{1+\beta
    p^2}E(p)
    \left(\hat{\phi}^{\dagger}(p)\hat{\phi}(p)+ \sp \mbox{const.}\right).
\end{eqnarray}
Annihilation and creation operators for one particle states have commutation relation:
\begin{eqnarray}
 \left[ \hat{\phi}(p) , \hat{\phi}^{\dagger}(p') \right]=(1+\beta p^2)\delta(p-p').
\end{eqnarray}
Canonical commutation relation is
\begin{eqnarray}
\left[\hat{\Phi}(p,t),\hat{\Pi}(p',t)\right]=i\hbar(1+\beta p^2)
\delta(p+p'). 
\end{eqnarray}
And Heisenberg equation is
\begin{eqnarray}
\frac{1}{i\hbar}\left[ \hat{\Phi}(p,t), {\cal H} \right]
=\frac{\partial}{\partial t}\hat{\Phi}(p,t) =\hat{\Pi}(p,t).
\end{eqnarray}
We find propagators from these quantity as 
\begin{eqnarray}
\langle T\hat{\Phi}(p,q)\hat{\Phi}(p',q')\rangle
&\equiv& \frac{1}{2\pi\hbar}
\int dtdt' \exp\left(-\frac{qt+q't'}{i\hbar}\right)\langle T\hat{\Phi}(p,t)\hat{\Phi}(p',t')\rangle \nonumber \\
&=&\frac{\hbar^3}{i} (1+\beta p^2)
  \delta(p+p')\delta(q+q')\frac{1}{-q^2+E(p)^2-i\epsilon}. \nonumber \\
\end{eqnarray}
On the other hand, the Lagrangian is 
\begin{eqnarray}
{\cal L} = -\frac{1}{2}\int^{\infty}_{-\infty}\frac{dp}{1+\beta
    p^2}\Phi(-p,t)\left[\partial_t^2+\left(\frac{E(p)}{\hbar}\right)^2\right]
\Phi(p,t). \label{finalLag}
\end{eqnarray}
And the path integral measure which is consistent with the canonical formalism is
\begin{eqnarray}
1= \int {\cal D}\Phi(p,t)\exp\left(-\frac{1}{2}\int dt dp \Phi(p,t)\Phi(-p,t)\right).
\end{eqnarray}

When we add same interaction terms to both Hamiltonian
(\ref{finalHam}) and Lagrangian (\ref{finalLag}) with opposite sign, it
is obvious that Feynman rules in both formalism are same form and then two theories which include the interactions represent the same
physics. 

It is easy to see that the quantization procedure developed above 
based on the path integral formalism does not have any difficulty 
in the generalization to higher dimensions. 
Lagrangian and path integral measure in $d+1$ dimensional spacetime can be written  
\begin{eqnarray}
{\cal L}&=& -\frac{1}{2}\int^{\infty}_{-\infty}d^dp(1+\beta
    \mbox{\boldmath$p$}^2)^{-1}\phi(-\mbox{\boldmath$p$},t)
    \left[\partial_t^2+\left(\frac{E(\mbox{\boldmath$p$})}{\hbar}\right)^2\right]
\phi(\mbox{\boldmath$p$},t), \\
1&=& \int {\cal D}\Phi(\mbox{\boldmath$p$},t)\exp\left(-\frac{1}{2}\int dtd^dp \Phi(\mbox{\boldmath$p$},t)\Phi(-\mbox{\boldmath$p$},t)\right).
\end{eqnarray}

We would like to make a comment here.
It might be seen that the algebra (\ref{GUP}) is a sort of cut off theory by a change of variables :
\begin{eqnarray}
\rho &\equiv&\frac{1}{\sqrt{\beta}}\tan^{-1}(\sqrt{\beta}p), \\
\left[ x , \rho \right]&=& i\hbar,
\end{eqnarray}
where
\begin{eqnarray}
-\frac{\pi}{2\sqrt{\beta}} < \rho < \frac{\pi}{2\sqrt{\beta}} .
\end{eqnarray}
However the theory described in this variable can not be regarded as a simple cut off theory of ordinary field theory. 
If we set a cut off $\frac{\pi}{2\sqrt{\beta}}$, the parameter $l$
appeared in the integration measure (\ref{measure}) must not take a
value 0 but 1 because of $d\rho=\frac{dp}{1+\beta p^2}$:
\begin{eqnarray}
1&=& \int {\cal D}\Phi(\rho,t)\exp\left(-\frac{1}{2}\int^{\frac{\pi}{2\sqrt{\beta}}}_{-\frac{\pi}{2\sqrt{\beta}}} dt d\rho \Phi(\rho,t)\Phi(-\rho,t)\right) \nonumber \\
 &=&\int {\cal D}\Phi(p,t)\exp\left(-\frac{1}{2}\int^{\infty}_{-\infty}
			       dt \frac{dp}{1+\beta p^2}
			       \Phi(p,t)\Phi(-p,t)\right).
\end{eqnarray}
Thus, our theory is not mere variable transformation of a ordinary quantum field theory. 
In other words, the method of quantizing fields considered in this paper 
is an another method to the well-known method of quantizing the
particle on finite interval space \cite{Ohnuki:1991uv}. 
In that paper, system has periodic boundary condition. 
The translation generator $\hat{p}$ is ill-defined on that space, instead the
operator $\exp(i\hat{p})$ is well-defined. 
In the same sense, the operator $\hat{\rho}$ is ill-defined and the operator
$\hat{p}\equiv \frac{1}{\sqrt{\beta}}\tan(\sqrt{\beta}\rho)$ is
well-defined in our formalism. 
\section{Summary}

In this paper we construct the scaler field theory in arbitrary
dimensions with path integral formalism based on generalized canonical 
commutation relations. 
We adopt the guideline that canonical formalism and path integral formalism are equivalent by using BJL prescription and a second
quantized Hamiltonian is formed by calculating the `expectation value' 
of first quantized Hamiltonian with a `wavefunction' replaced by 
the quantized field \cite{Weinberg:1995mt}. We first consider the deformation in 1+1 dimensional canonical formalism, then we construct path integral formalism by BJL
prescription. In the path integral formalism there is not any difficulty in
extension to higher dimension.

\begin{center} \begin{large}
Acknowledgments
\end{large} \end{center}

We would like to thank K. Fujikawa and M. Hayakawa for fruitful discussions.
We also thank T. Tada for careful reading of the manuscript. 
One of the author (T. M.) also thanks Y. Kimura, S. Matsuura, 
S. Seki and M. Tachibana for valuable discussions and comments. 
This work is supported by Special Postdoctoral Researchers Program at RIKEN.

\appendix


\end{document}